\documentclass[slac_one]{revtex4}
\usepackage{graphicx}
\usepackage{fancyhdr}
\usepackage{epsfig}

\def\et  {\ensuremath{E_{T}}\xspace}

\input{babarsym}

\begin{document}

\title{{\small{Hadron Collider Physics Symposium (HCP2008),
Galena, Illinois, USA}}\\ 
\vspace{12pt}
Commissioning of Particle ID at ATLAS and CMS with Early LHC Data} 

%

\author{T. Berger-Hryn'ova}
\affiliation{CERN, 1211 Geneve 23, Switzerland}
\author{on behalf of the ATLAS and CMS collaborations}

\begin{abstract}
This paper describes latest results on lepton (electron, muon and tau) 
and photon particle identification at the ATLAS and CMS experiments, 
with emphasis on how the particle identification can be validated and its 
performance determined using early LHC data. 
\end{abstract}

\maketitle

\thispagestyle{fancy}


\section{INTRODUCTION} 

ATLAS and CMS are the two general-purpose experiments, which have recently completed installation at
the CERN Large Hadron Collider (LHC). Many physics processes which are 
currently out of reach at existing colliders will become accessible at the LHC.
These processes range from the production of scalar Higgs bosons or of new vector gauge bosons to that 
of supersymmetric particles or of TeV-scale resonances
resulting from presence of large extra-dimensions.  
Charged leptons often provide distinctive signatures for such new processes, but they also 
appear in the final state of many standard-model (SM) processes involving electroweak (EW)
bosons or top quarks. These SM processes constitute the dominant background to new signals and therefore need to be well 
understood. They will also be used as calibration processes to understand in detail the performance of the detector.  

Thus, lepton (and photon) identification is very important for the physics program of 
ATLAS and CMS. 
With a bunch-crossing rate of 40$\,$MHz at design luminosity, of which only 
200-300$\,$Hz are planned to be recorded to mass storage, particle identification is 
extremely important also at the trigger level and it will be crucial 
to commission it as early as possible. 

The main data samples used to understand the detector performance and 
the lepton identification are well-known EW processes, such as $Z\to ll$ and 
$W\to l\nu$. They provide important benchmark channels for calibration, 
alignment and monitoring of the detector performance. 
However, at the expected start-up luminosities of $10^{31}-10^{32}$\cms, the total 
bandwidth for EW processes will be below 1$\,$Hz, compared to a few tens of Hz 
expected at $10^{33}$\cms, which will severely limit the statistics of such  
samples. On the other hand, at these lower luminosities, the triggers can be operated with lower 
\et-thresholds than those planned for the LHC design luminosity. This would potentially give access to 
other more abundant data samples, such as direct $J/\psi$, $\Upsilon$, etc. 

This paper describes the commissioning strategies in the following order: 
identifications of photons, electrons, muons and $\tau-$leptons.
For a detailed description of the ATLAS and CMS detectors and of their current commissioning status please 
see Refs.~\cite{thomas,atldet,luca}.

\section{Photons and Electrons}

Photon and electron identification in ATLAS and CMS is extremely challenging: the electron 
to jet ratio is $\approx 10^{-5}$ at 40\gev. Photon and electron reconstruction in both 
experiments starts by the detection of clusters in the electromagnetic calorimeters. 
Electron candidates are also required to have an inner-detector track loosely matched to the cluster. 
In ATLAS, there is also a 
dedicated low-\pt electron reconstruction algorithm, which extrapolates inner-detector
tracks to the EM calorimeter. 
The reconstruction of photons and electrons is quite challenging in both 
experiments because of the $\approx 0.4-2.4 X_0$ of material in the inner 
detectors. For example, in ATLAS, electrons with \et$=$25\gev lose on average $30-60\%$ 
of their energy before reaching the EM calorimeter, and 
between $10$ and $60\%$ of photons convert into an electron-positron pair.
For a more detailed description of the electron and photon reconstruction 
in ATLAS and CMS, see Ref.~\cite{charlot}.

Despite some significant differences in detector technologies and intrinsic performance 
(e.g. the sampling LAr EM calorimeter of ATLAS and the homogeneous PbWO$_4$ crystal EM calorimeter of CMS), 
the performances expected for electrons and photons are similar for both experiments
in terms of efficiency and accuracy. For example 
electrons of 50\gev energy are expected to be measured with a $1.5-2.5\%$ energy resolution in ATLAS
and with a 2$\%$ energy resolution in CMS. 
Photons with a 100\gev energy are expected to be measured with a $1.0-1.5\%$ 
energy resolution in ATLAS. For $70\%$ of the photons which do not convert too early in the tracker, 
the energy resolution is expected to be better in CMS, e.g. $\approx 0.8\%$ due to the superior intrinsic
accuracy of the crystal calorimeter. 

Very high rejection of the large background from hadronic jets is another important aspect of 
electron and photon identification at the LHC. In ATLAS, 
for a photon reconstruction efficiency of $85\%$, one can obtain a jet rejection 
of $\approx 10000$ averaged over all jets types (quark and gluon jets).
For tight electron selection and for \et$>20$\gev, 
the identification efficiency of isolated electrons from $Z$ decays is $65\%$ (it drops to $\approx 25\%$
for non-isolated electrons from $b/c-$decays) with a jet rejection of 
$\approx 10^5$. After such tight electron selection of inclusive electrons with \et$>$20\gev
the remaining sample 
is expected to contain $~10\%$ of isolated and $~65\%$ of non-isolated 
electrons, with the remaining $~25\%$ expected to consist of residual background from hadronic jets dominated
by charged hadrons. The use of multivariate 
methods improves the rejection by $~50\%$ for a fixed efficiency or improves 
the efficiency by $5-10\%$ for a fixed rejection. 

Obviously, those expected performances will have to be validated with real data. 
Both experiments are planning to use 
the early data to understand the tracker (alignment, material distribution), to perform detailed intrinsic calibrations
of the EM calorimeters, as well as 
to measure the trigger efficiency. All these aspects are crucial to obtain a detailed understanding of the 
electron and photon identification performances. 

\subsection{Photons}

Photon identification is particularly challenging, since there do not exist 
any prominent di-photon resonances at high mass which could be 
used for calibration and efficiency measurement purposes. Thus, the understanding of photon identification  
in the early data will have to rely heavily on 
the understanding of electrons, using simulations to account for the 
small differences between them. As soon as a few \invfb of data become available 
$Z\to ee\gamma$ and $Z\to \mu\mu\gamma$ decays will be used to improve the understanding of high-\pt
photons further. 
In CMS the achievable statistical error on the photon efficiency is estimated 
to be $0.1\%$ with 1\invfb of data.  
At much lower \pt, one could use $\pi^0$ and $\eta$ decays to $\gamma\gamma$ 
to understand photon identification in early data. 
Converted photons 
from $\pi^0/\eta$ decays will be extremely useful to obtain a detailed mapping of 
the tracker material. This is strongly correlated with a precise 
EM calorimeter inter-calibration, as discussed below. 

\subsection{Electrons}\label{el}
A detailed understanding of electron reconstruction and identification at the LHC will rely strongly 
on the presence of 
``standard candles'', such as the Z and W resonances, for the calibration of the energy scale,
for the EM inter-calibration, and for a precise understanding of the detector and trigger efficiencies, etc. 

\begin{figure}[h!]
\centering
{\includegraphics[width=0.40\linewidth]{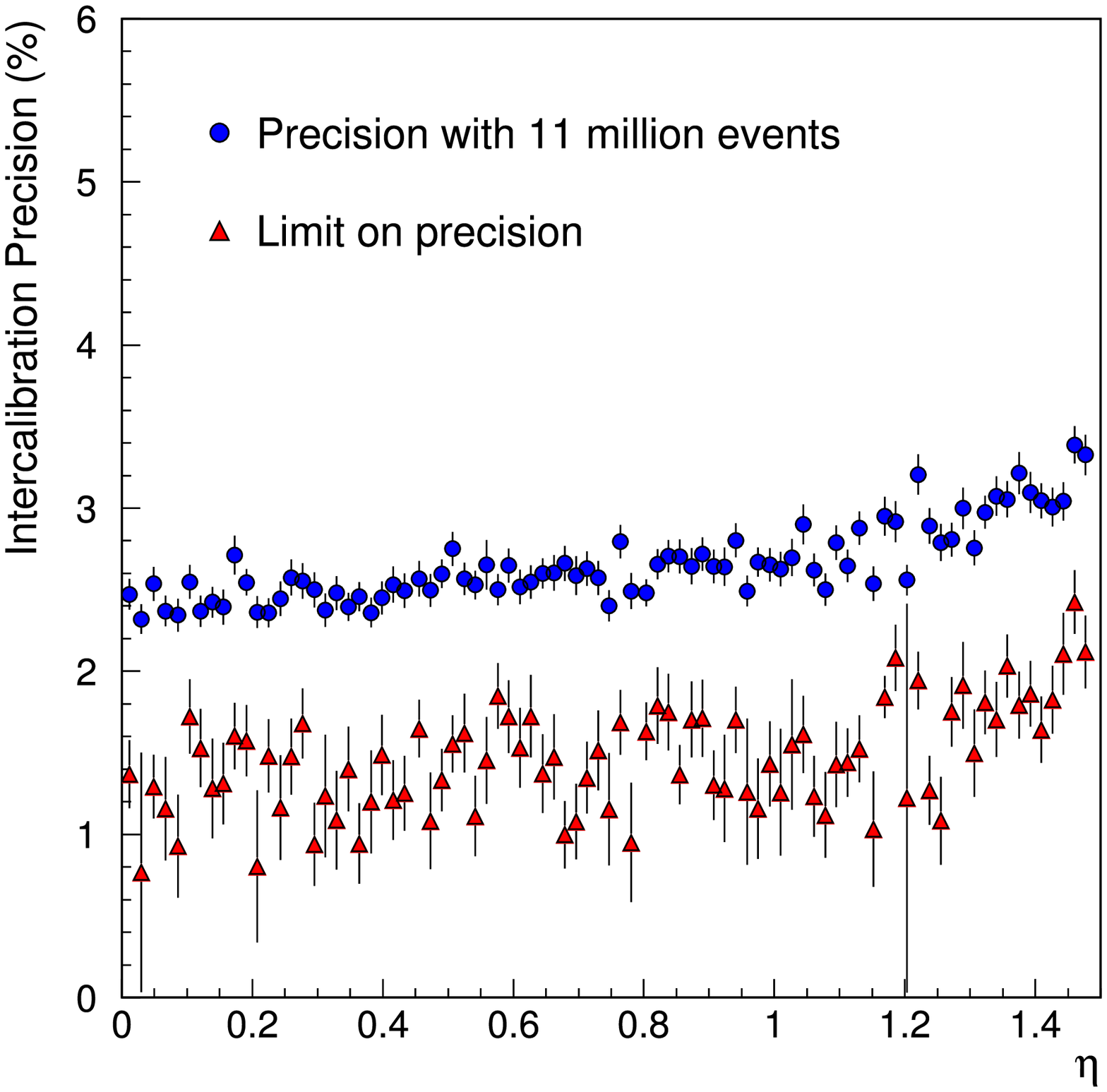}}
{\includegraphics[width=0.40\linewidth]{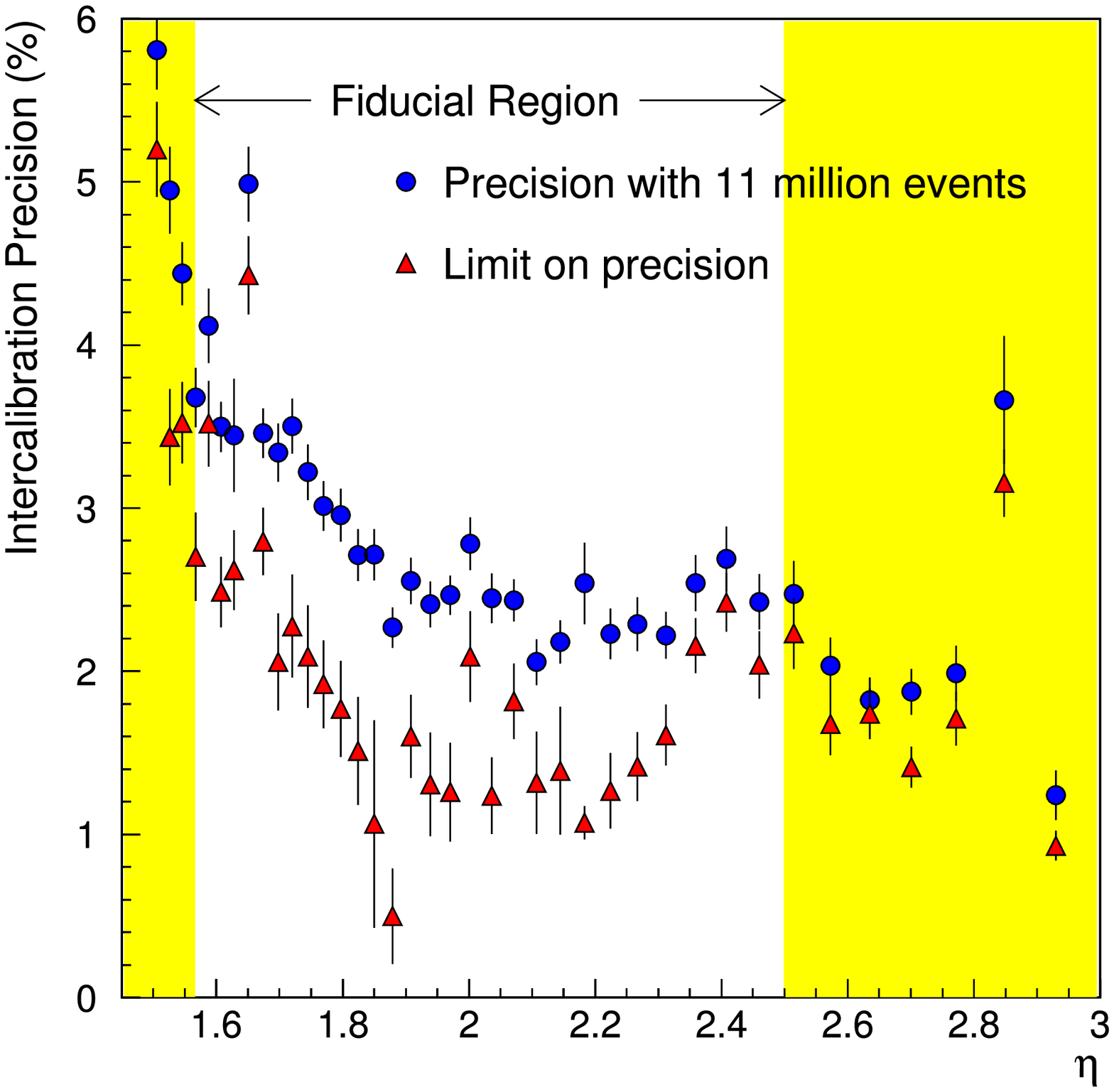}}
\caption{CMS inter-calibration precision in barrel (left) and endcap (right) which can be 
obtained with 11 million Level-1 jet trigger events and the limit on the inter-calibration 
precision due to tracker material inhomogeneity as a function of $\eta$ (from Ref.~\cite{cmstdr}).} \label{jets}
\end{figure}

Both experiments are planning to use the precise knowledge of the $Z$ mass
to perform an accurate EM calorimeter inter-calibration. Since the ATLAS EM calorimeter is locally very uniform by 
construction, only an inter-calibration of large regions is required. 
The initial spread from region to region is conservatively assumed to be
approximately $1.5-2\%$. Provided the material distribution 
in front of the EM calorimeter is well-understood, the specified inter-calibration precision of 0.7$\%$ 
between regions of 0.2$\times$0.4 in $\eta\times\phi$
can be achieved in the case of ATLAS with 100\invpb of data. 
In the CMS EM calorimeter, the energy response may vary with a spread of $~3-4\%$ 
from crystal to crystal, and therefore a precise 
inter-crystal calibration has to be performed. Initially, single-jet triggers will be used to 
inter-calibrate $\phi$ rings at fixed $\eta$ with an accuracy of a few percent, as is shown in Fig.~\ref{jets}. 
For the final precision the $Z\to ee$ sample corresponding to $\approx$2\invfb of data 
is needed to reach the specified target of 0.5$\%$. 

\begin{figure}[ht!]
\centering
{\includegraphics[width=0.40\linewidth]{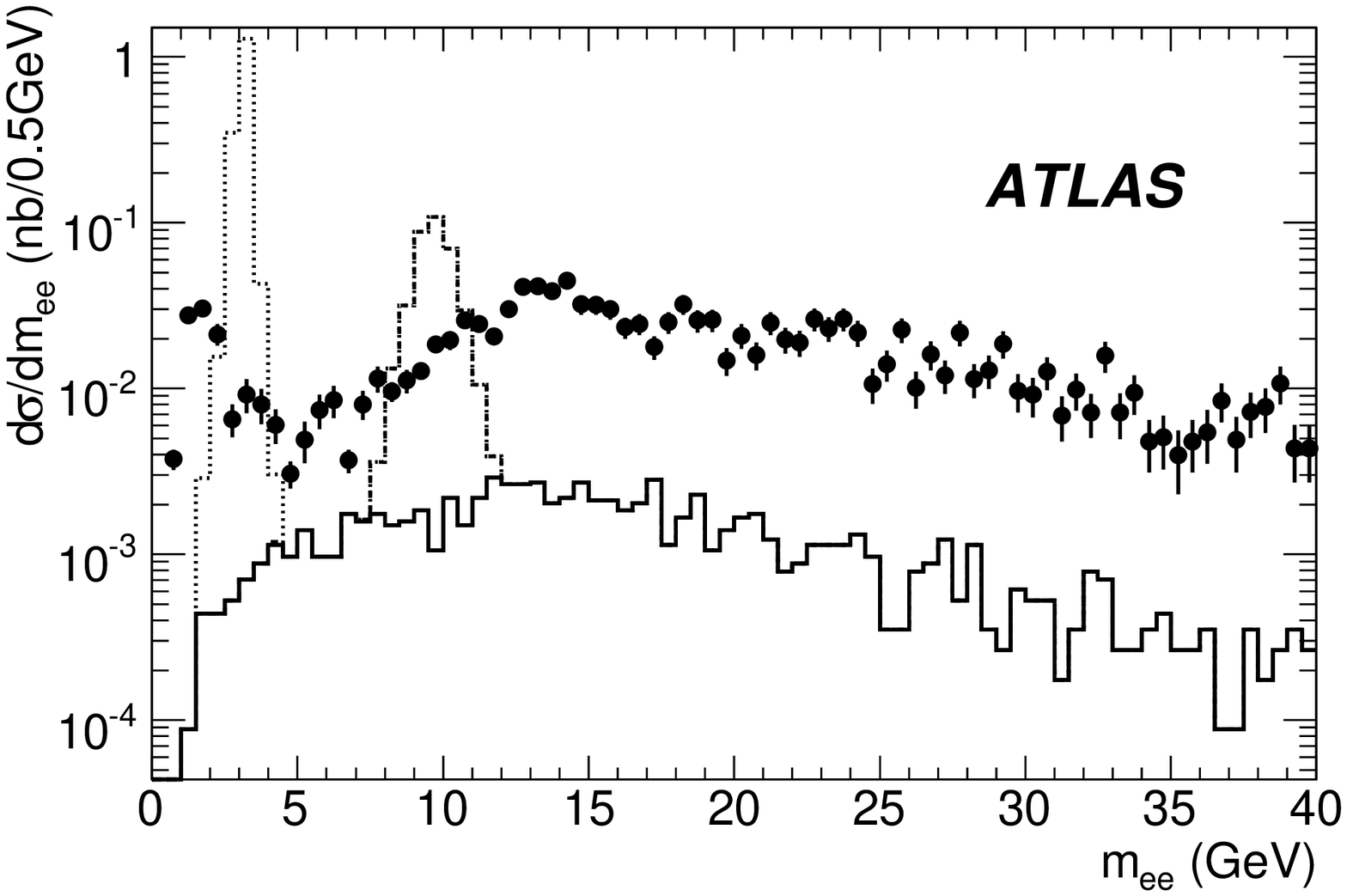}}
{\includegraphics[width=0.40\linewidth]{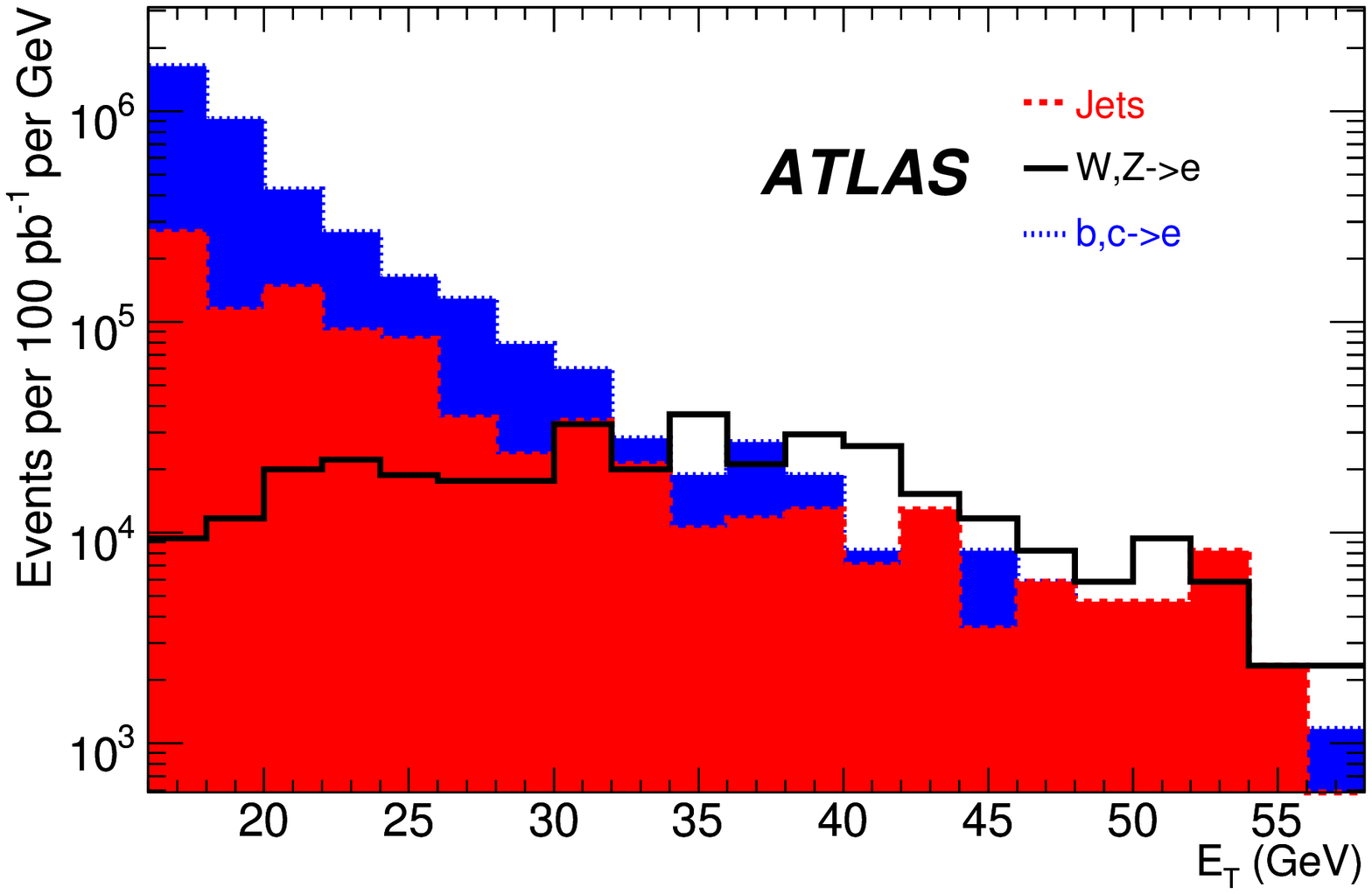}}
\caption{(Left) Expected differential cross-section for low-mass electron pairs using 
the dedicated ATLAS low-\pt dielectron trigger. Shown is the invariant di-electron mass 
distribution reconstructed using tracks for $J/\psi \to e e$ decays (dotted histogram), 
$\Upsilon \to e e$ decays (dashed histogram) and Drell-Yan production (full histogram). 
Also shown is the expected background (full circles).
(Right) Expected differential rates in ATLAS as a function of electron \et for 
100\invpb of data and for electrons satisfying tight single-electron selection.
Shown are the isolated electrons from $W$ and $Z$ decays (solid histogram),
the non-isolated electrons from $b,c\to e$ decays (light gray/red histogram), and the 
estimated QCD background (dark gray/blue histogram).} \label{electron}
\end{figure}

\begin{figure}[ht!]
\centering
{\includegraphics[width=0.40\linewidth]{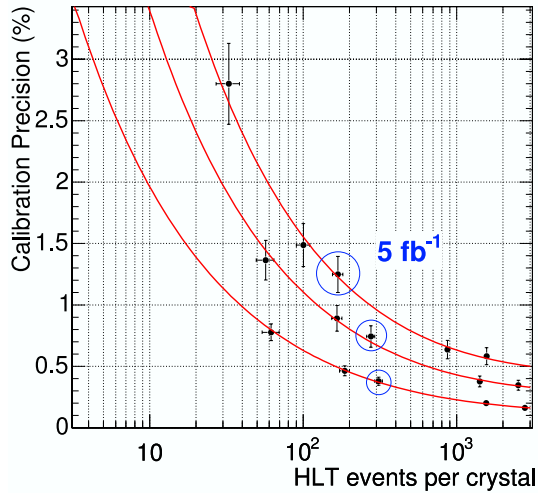}}
\caption{CMS inter-calibration precision as a function of HLT events per crystal for different $\eta$
regions. Upper curve: last 10 crystals in the EM calorimeter barrel ($1.305<\eta<1.479$), 
middle curve: 10 crystals in the middle EM calorimeter barre ($0.783<|\eta|<0.957$), lower curve:
the first 15 crystals in the EM calorimeter barrel ($0.0<|\eta|<0.261$). The third point on each line
gives the precision for 5\invfb of integrated luminosity (from Ref.~\cite{cmstdr}).} \label{ew}
\end{figure}

In ATLAS, specific triggers have been implemented for electron pairs from direct $J/\psi$ and $\Upsilon$ production,
which are expected to provide about 200k and 50k events respectively for 100\invpb of data (compared to 
50k of events on the $Z$ resonance). The expected signal and background dielectron mass distributions after tight
electron selection are shown in Fig.~\ref{electron} (left). 
The obtained $J/\psi$ sample can also be used for the calorimeter inter-calibration and 
preliminary studies~\cite{fred} show that a precision of $\approx 0.6\%$ may be achieved.   
In addition to a cross-check on the inter-calibration result obtained using the $Z$ resonance, 
the $J/\psi$ and $\Upsilon$ samples will also provide an
in-situ check of the linearity of the EM calorimeter response as a function 
of electron energy. The additional electron identification 
capabilities provided by the ATLAS transition radiation detector may turn out to be 
crucial to identify low-energy electrons
from $J/\psi$ and $\Upsilon$ decays, since they are usually not as well isolated as those 
from $Z$ decay and since the charged-pion combinatorial background is very large. 

Isolated single electrons from $W\to e\nu$ decays will provide further information for 
tracker-EM calorimeter inter-calibration by comparing the energy deposited in the calorimeter to 
the track momentum measurement. As shown in Fig.~\ref{ew}, an inter-calibration precision of 
$0.5-1.5\%$ may be achieved in CMS with 5\invfb of data. Non-isolated electrons from $b,c\to e$ decays could 
also be used for this purpose during the early data-taking while the single-electron trigger 
thresholds are low. As one can see from Fig.~\ref{electron} (right), these decays are 
dominant for \et$<$35\gev and they should provide ten times more electrons with \et$>$10\gev 
than those from $W$ decay.

Since the trigger efficiency represents a fundamental ingredient of any physics analysis, 
it must be verified as independently as possible from simulations and as a function of the LHC physics and detector
operation conditions. One of the most widely spread techniques under study is the 
``tag-and-probe'' method, where an event is triggered by one of the electrons in a $Z\to ee$ decay and 
the efficiency to trigger the other electron is measured using offline information. 
With 100\invpb of data, one will measure the trigger efficiency for electrons of $\et>20\gev$
with a statistical accuracy of approximately 0.2$\%$~\cite{atldet,cmstag}. 

\section{Muons}

Muons are identified through matching reconstructed tracks in 
the inner tracker and the muon spectrometer.
ATLAS and CMS follow complementary concepts of muon detection. ATLAS has an 
instrumented air-toroid magnetic system serving as a stand-alone high-precision muon spectrometer. CMS relies
on high bending power and momentum resolution in the inner tracker, and on the iron yoke used for the return of 
its solenoidal magnetic field. The iron yoke is instrumented with chambers used for muon identification and triggering.
Stand-alone muon tracks can be reconstructed, but in order to obtain
a precise momentum measurement it is essential to combine the
inner-tracker information with that from the muon chambers.

\begin{figure}
\centering
{\includegraphics[width=0.60\linewidth]{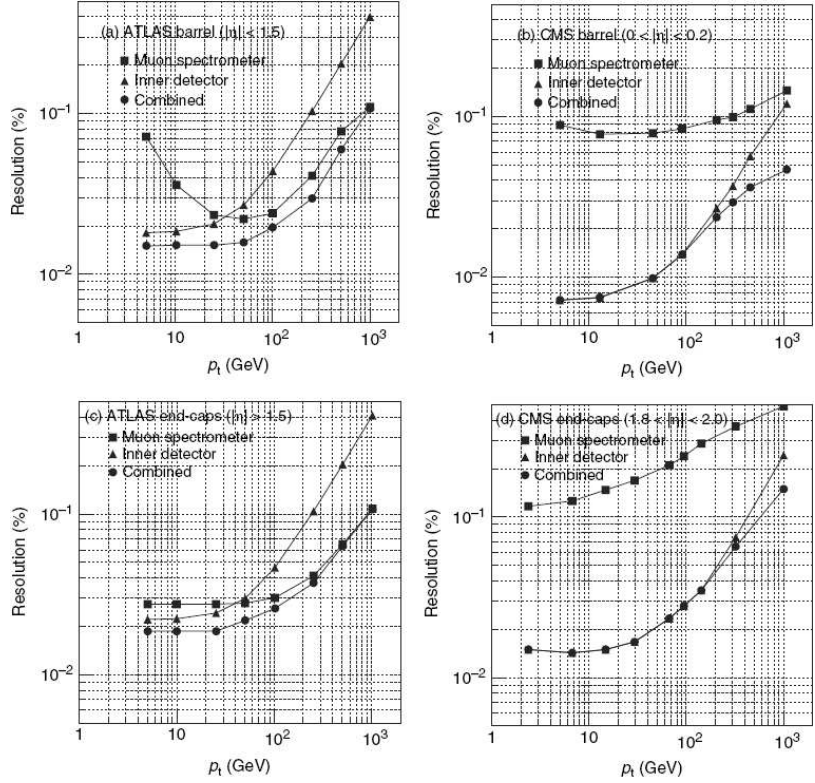}}
\caption{Relative momentum resolution as a function of the muon transverse momentum 
showing the stand-alone resolution of muon systems, the stand-alone resolution of the inner tracker and 
the combined resolution: (a) and (c) in ATLAS for $|\eta|<1.5$ and $|\eta|>1.5$, respectively; (b) and 
(d) in CMS for $0<|\eta|<0.2$ and $1.8<|\eta|<2.0$, respectively. From Ref.~\cite{gigi}.} \label{muonres}
\end{figure}

\begin{figure}
\centering
{\includegraphics[width=0.50\linewidth]{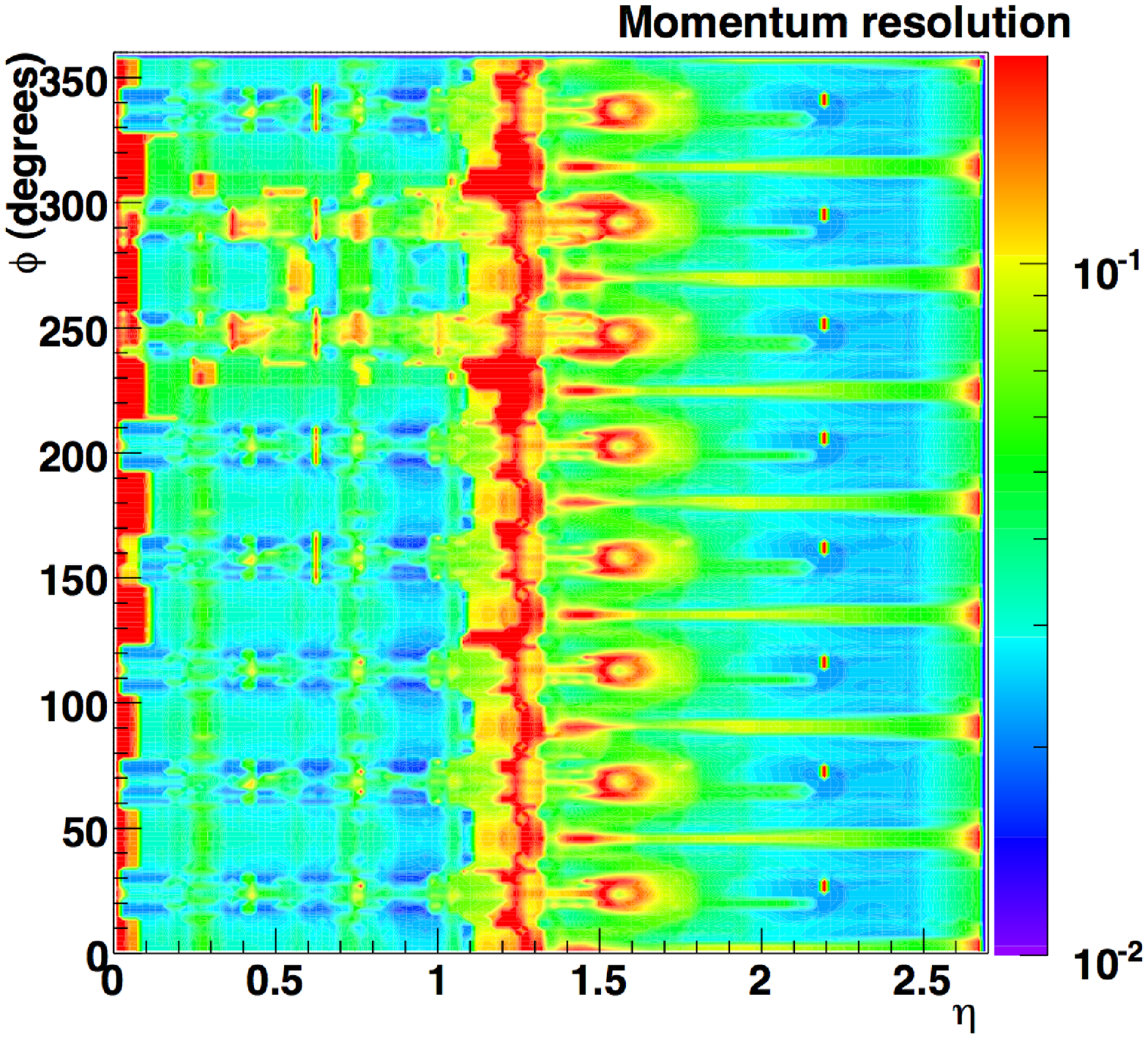}}{\includegraphics[trim=11cm 0cm 0cm 0cm,clip,width=0.4\linewidth]{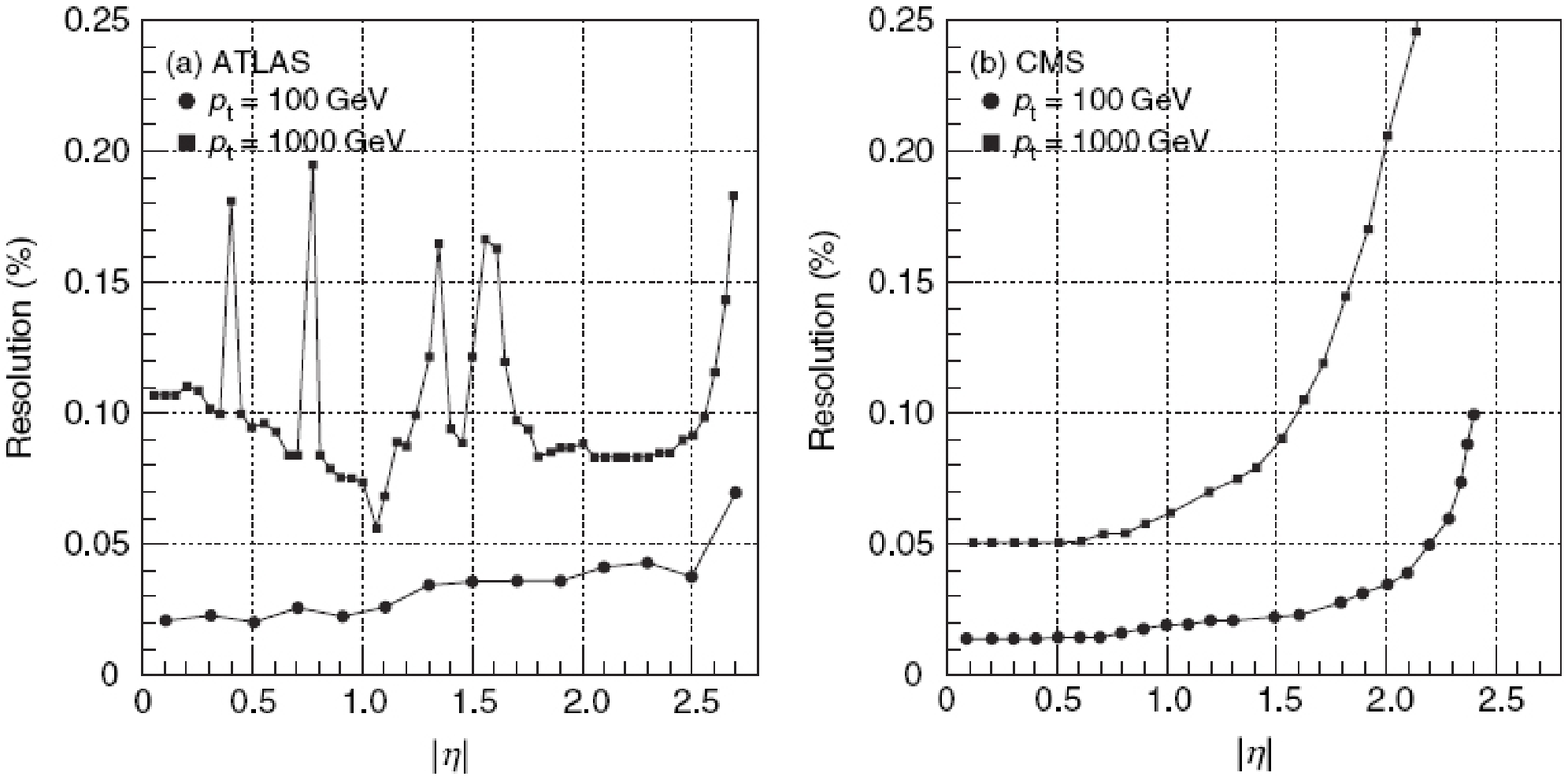}}
\caption{(left) For muons with \pt$=$100\gev, expected fractional stand-alone momentum resolution in ATLAS
as a function of $\phi$ and $|\eta|$ (from Ref.~\cite{atldet}). No momentum measurement is possible at $|\eta|<$0.1 over 
most of the azimuth, nor at $|\eta|=1.3$ because of holes in the acceptance 
of the muon spectrometer. 
(right) Relative momentum resolution for combined muon reconstruction 
in CMS as a function of $\eta$ for two
different values of transverse momentum (from Ref.~\cite{gigi}).} \label{muoneff}
\end{figure}

The momentum resolution of the ATLAS and CMS detectors are shown in Fig.~\ref{muonres}.
Whereas the stand-alone muon resolution is fairly uniform over most of the 
$\eta-\phi$ plane in ATLAS as is shown in Fig.~\ref{muoneff}(left) 
with optimal resolution achieved at 100\gev, the resolution degrades at 
high $\eta$ in CMS because of the limited coverage of the solenoid (Fig.~\ref{muoneff}(right)). 
The resolution of the combined measurement in the barrel region is slightly 
better in CMS owing to the higher precision of the measurement in the tracking system, 
whereas the reverse is true in the end-cap region owing to the better coverage of the ATLAS
muon spectrometer at higher rapidities.  

The muon reconstruction efficiencies are high (above $96\%$) for both experiments with fake rates of  
about $1\%$ from neutron background at the LHC design luminosity, see Ref.~\cite{kortner} for details. 
The combination of inner tracker, calorimeter and muon spectrometer measurements is required in ATLAS 
to minimize inefficiencies due to holes in the coverage of the muon spectrometer 
system as is shown in Fig.~\ref{muoneff}.

\begin{figure}
\centering
{\includegraphics[width=0.40\linewidth]{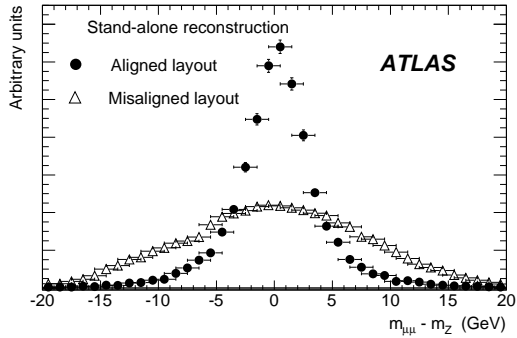}}
\caption{Difference between reconstructed and true dimuon mass from $Z\to\mu\mu$ decays, as obtained from 
an aligned and a misaligned muon spectrometer in ATLAS. 
} \label{muon1}
\end{figure}

\begin{figure}
\centering
{\includegraphics[width=0.40\linewidth]{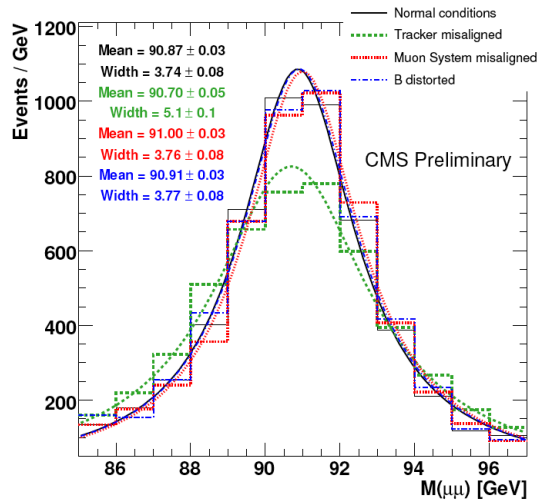}}
\caption{Effect on the reconstructed $Z$ boson mass in CMS 
of the misalignment and of the uncertainties on the 
magnetic field after the muon scale calibration. Fits are sums of a Lorentzian and a 
linear function. See Ref.~\cite{cmsmu} for details.
} \label{muon2}
\end{figure}

\begin{figure}
\centering
{\includegraphics[width=0.40\linewidth,height=6cm]{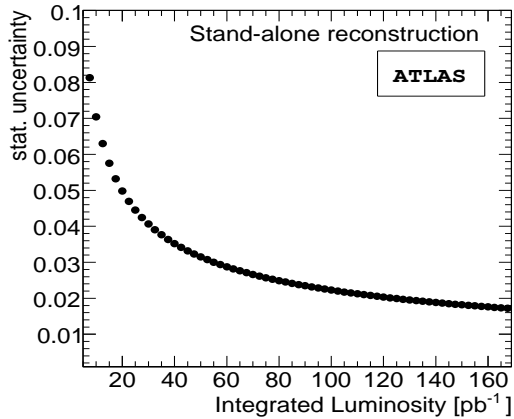}}
\caption{
Average statistical error expected for the measured muon reconstruction efficiency in ATLAS for 320 regions 
of the muon spectrometer as a function of integrated luminosity.} \label{muon3}
\end{figure}

The large expected rates of $Z\to\mu\mu$ decays provide an excellent tool to untangle various effects
which might lead to distortions of the measured dimuon invariant mass spectrum. One example is shown 
in Fig.~\ref{muon1} for ATLAS stand-alone muon measurements, where the performance obtained with a misaligned 
layout (with random displacements of 1\mm and a random rotations of 1\mrad of each chamber in the muon spectrometer)  
is compared to that expected from a perfectly aligned layout. The large assumed misalignments lead to an increase in the 
fitted Gaussian resolution of the dimuon reconstructed invariant mass from 2.5\gev to 8\gev.
Similar distortions including a potential shift in the mean of the peak 
could be caused by muon momentum scale biases and uncertainties in the 
initial knowledge of the magnetic field as is shown in Fig.~\ref{muon2}. 

The muon reconstruction and identification efficiency will also be measured from data using 
$Z\to\mu\mu$ decays and the tag-and-probe method described in Section~\ref{el}, with similar results 
expected in terms of the accuracy of the measurement. This type of analysis can be extended to 
study efficiency with increasing granularity as the integrated luminosity increases, 
in particular as a function of $\eta$ and $\phi$. 
An example of the expected statistical uncertainty on the reconstruction efficiency 
in the muon spectrometer, 
averaged over 320 regions in the $\eta-\phi$ plane, is shown in Fig.~\ref{muon3}. 
Other resonances, such as $J/\psi$ and $\Upsilon$, could also be used for this analysis
at lower initial luminosities. The expected rates for 100\invpb of data in ATLAS/CMS are 1.6$\,$M $J/\psi$, 
300k $\Upsilon$ and 60k $Z$-decays.

\section{$\tau-$leptons}

The $\tau-$lepton is the heaviest lepton observed to-date and it decays to electrons, muons and hadrons.
This section concentrates on the identification on $\tau-$leptons decaying hadronically. They  
are reconstructed by matching 
narrow calorimeter clusters with a small number of tracks reconstructed in the inner tracker. 

\begin{figure}[ht!]
\centering
{\includegraphics[width=0.40\linewidth]{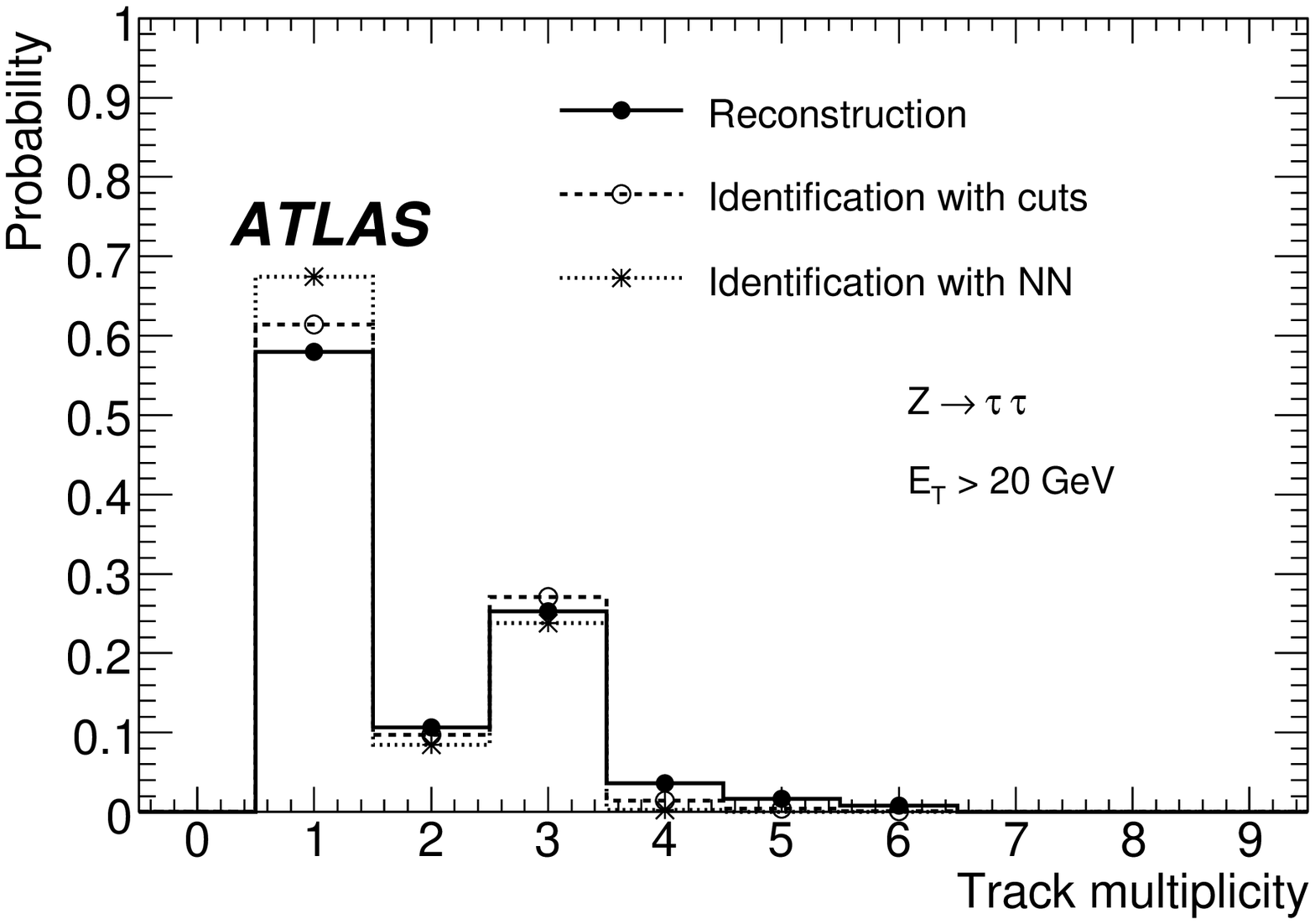}}
{\includegraphics[width=0.40\linewidth]{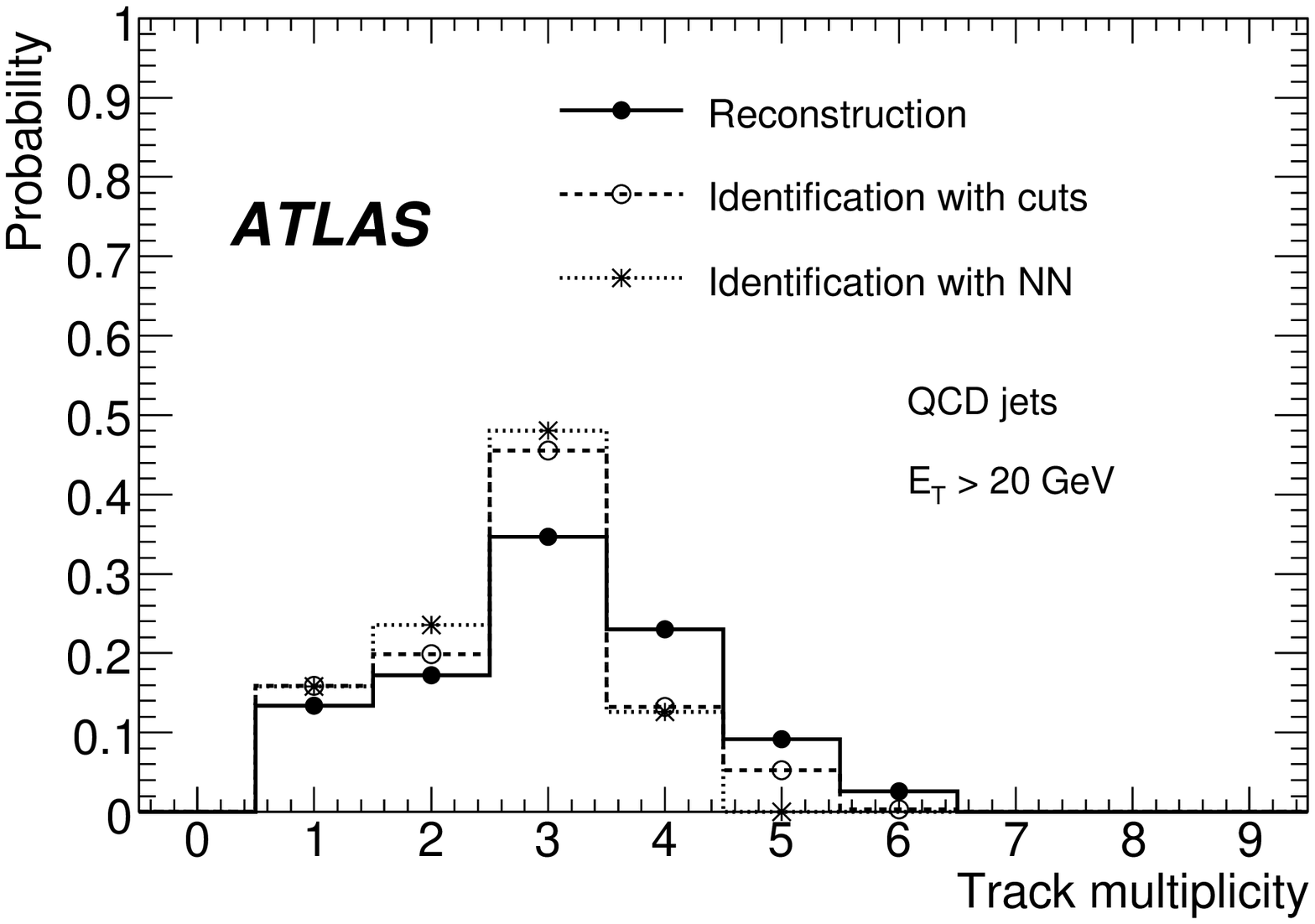}}
\caption{Track multiplicity distributions expected in ATLAS for hadronic $\tau$-decays (left) and
for the background from hadronic jets (right), for $\tau-$candidates   
with visible transverse energy above 20\gev reconstructed using a track-based $\tau$-identification 
algorithm. The distributions are shown after reconstruction, after cut-based 
identification, and finally after applying a neural network discrimination 
technique, resulting in an efficiency of 30$\%$ for the signal.
} \label{tau}
\end{figure}

One will need to demonstrate that the observed signals from $W\to\tau\nu$ and $Z\to\tau\tau$ decays are in  
agreement with fundamental properties of the $\tau-$lepton (e.g. visible mass and decay length 
measurement for three-prong decays).
One of the distinctive features of hadronic $\tau$ decays is the track multiplicity spectrum shown in Fig.~\ref{tau} 
for $Z\to\tau\tau$ decays (left) and for jets (right). The distributions are shown after the reconstruction 
step, after a cut-based identification algorithm and finally after applying a multi-variate 
discrimination technique using a neural network. Figure~\ref{tau} shows that the signal track 
multiplicity distributions do not depend strongly on the reconstruction and identification cuts used. It 
also shows that the track multiplicity in the 
QCD jet sample is quite different from that in the signal sample
and that the candidates with track multiplicity above three may be used to 
normalize the background. This would allow a reasonably precise calibration of the performance
of the $\tau$-identification algorithms using real data, provided the rejection against QCD jets
is proven to be sufficient to extract clean signals in the single-prong and three-prong categories.  

Triggering on hadronic $\tau-$decays is extremely challenging because of the huge backgrounds from 
hadronic jets and of the requirement to bias minimally the track multiplicity spectrum. 
The $\tau-$triggers increase the discovery potential for many physics channels and 
in rare cases provide the only trigger, e.g. $W\to\tau\nu$ decays.

$W\to\tau\nu$ decays provide the most abundant source of $\tau-$leptons in SM processes. 
Due to trigger rate limitations, they are only expected to be accessible at initial low luminosities. 
The dominant background for it arises from hardonic jets and the signal-to-background ratio is expected to be 
a factor of ten worse than that 
observed at the TeVatron. Preliminary studies show that it is possible to reach a signal-to-background
ratio of about three in this channel with $\approx 1500$ $W\to\tau\nu$ signal events for 100\invpb of data. 
The goal of such an analysis would be to show track-multiplicity spectrum of identified $\tau-$leptons as a proof
of $\tau-$lepton observability. 

The most interesting SM sample is that from $Z\to\tau\tau$ decays. Although the expected rate is
ten times less than $W\to\tau\nu$ decays, this process
exhibits a more interesting topology and is easily triggered by requiring a single lepton. 
Same-sign events which are nearly signal-free will be used to control the 
dominant QCD background in the signal-enriched
opposite-sign events. The goal of this analysis is to set $\tau$ energy scale from the excess of 
signal events in the 
invariant mass of the visible decay products. The complete $Z$ invariant mass can be also reconstructed 
in the collinear approximation. Preliminary studies from ATLAS expect to see about 520 signal events and 
85 background events in 100\invpb. The visible decay mass could be used to determine the $\tau$ 
energy scale with a precision of 3$\%$ (taking into account only statistical errors). Also assuming 
$\tau$s are well-calibrated and using additional selection ($\approx$ 200 signal and 20 background
events remaining) and the collinear approximation, one could determine $E_T^{miss}$ scale with $3\%$ 
statistical precision. One should note that the $E_T^{miss}$ scale could be more easily 
understood with $W\to e(\mu)\nu$ events. 

\section{Acknowledgments}

The author would like to thank Daniel Froidevaux, Pascal Vanlaer, Oliver Kortner, Martijn Mulders and 
Elzbieta Richter-Was for the fruitful discussions and exchanges during this work.

\end{document}